\documentclass[10pt,letterpaper]{article}
\usepackage{opex3}
\usepackage{cite}

\newcommand\pictc[5]{\begin{figure}[htbp]
                          \centering
                       \includegraphics*[width=#1\columnwidth,]{#3}
                   \protect\caption{\protect\label{fig:#4} #5}
                    \end{figure}            }
\newcommand\pict[4][1]{\pictc{#1}{!tb}{#2}{#3}{#4}}
\newcommand\rpict[1]{\ref{fig:#1}}

\newcommand\leqt[1]{\protect\label{eq:#1}}
\newcommand\reqtn[1]{\ref{eq:#1}}
\newcommand\reqt[1]{(\reqtn{#1})}

\begin{document}
\begin{sloppy}

\title{Soliton control in modulated optically-induced photonic lattices}

\author{Ivan L. Garanovich, Andrey A. Sukhorukov and Yuri S. Kivshar}

\address{\mbox{Nonlinear~Physics~Centre~and~Centre~for Ultra-high bandwidth~Devices~for~Optical~Systems~(CUDOS)},
Research~School~of~Physical~Sciences~and~Engineering, Australian~National~University,
Canberra, ACT 0200, Australia}

\email{ilg124@rsphysse.anu.edu.au}
\homepage{http://www.rsphysse.anu.edu.au/nonlinear} 

\begin{abstract}
We discuss soliton control in reconfigurable optically-induced photonic
lattices created by three interfering beams. We reveal novel
dynamical regimes for strongly localized solitons, including
binary switching and soliton revivals through
resonant wave mixing.
\end{abstract}

\ocis{(190.0190) Nonlinear optics; 
      (190.4420) Nonlinear optics, transverse effects  in; 
      (190.5330) Photorefractive nonlinear optics.}


\section{Introduction}

The study of nonlinear light propagation in periodic photonic
lattices has attracted a strong interest due to many possibilities
of the light control offered by an interplay between the effects
of nonlinearity and periodicity. In particular, a periodic
modulation of the refractive index modifies substantially both the
linear spectrum and wave
diffraction~\cite{Russell:1995-585:ConfinedElectrons} and,
consequently, strongly affects the nonlinear propagation and
localization of light in the form of optical
solitons~\cite{Kivshar:2003:OpticalSolitons}.

Recent theoretical and experimental studies have demonstrated nonlinear localization of light in the optically-induced photonic lattices where the refractive index is modulated periodically in the transverse direction by an interference pattern of plane waves that illuminate a photorefractive crystal with a strong electro-optic anisotropy~\cite{Efremidis:2002-46602:PRE, Fleischer:2003-23902:PRL, Fleischer:2003-147:NAT, Neshev:2003-710:OL}. When the lattice-forming waves are polarized orthogonally to the {\em c}-axis of the photorefractive crystal, the periodic interference pattern propagates in the diffraction-free linear regime, thus creating a refractive-index modulation similar to that in weakly coupled waveguide array structures~\cite{Christodoulides:2003-817:NAT}.
Such optically-induced one-dimensional photonic lattices have been employed to demonstrate many fundamental concepts of the linear and nonlinear light propagation in periodic photonic systems, including the generation of lattice~\cite{Fleischer:2003-23902:PRL, Neshev:2003-710:OL, Martin:2004-123902:PRL} and spatial gap solitons in defocusing~\cite{Fleischer:2003-23902:PRL, Fleischer:2003-147:NAT} and self-focusing~\cite{Neshev:2004-83905:PRL} regimes, Bragg scattering and Bloch-wave steering~\cite{Sukhorukov:2004-93901:PRL}, tunable negative refraction~\cite{Rosberg:physics/0503226:ARXIV}, etc.

In this work, we study the soliton propagation in dynamic optical
lattices and identify novel effects associated with the
optically-induced refractive index modulation in the longitudinal
direction. Such lattices can be created by several interfering
beams, which are inclined at different angles with respect to the
crystal. In particular, we consider modulated photonic
lattices, created in a photorefractive nonlinear medium by 
{\em three interfering beams}, as shown in the examples
presented in Figs.~\rpict{potentials}(a-c). Here $z$ is the propagation coordinate
and beams experience one-dimensional diffraction only along the $x$~direction.

\pict{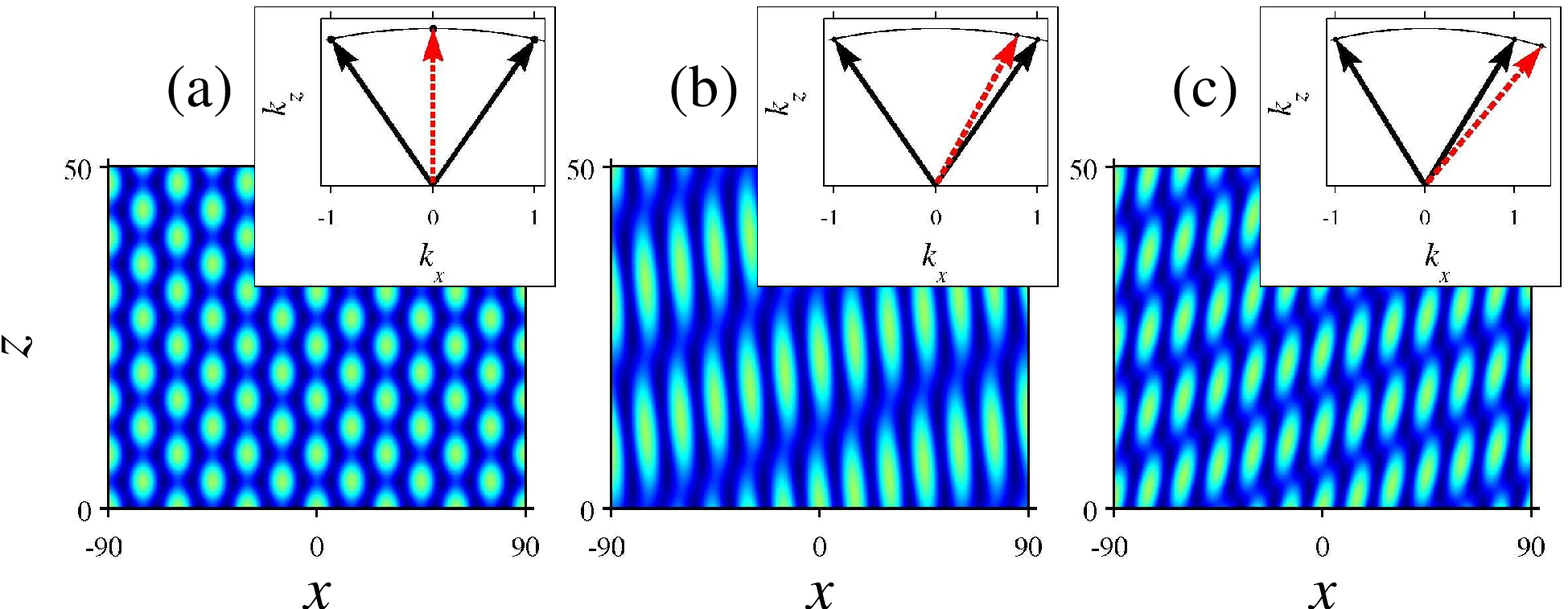}{potentials}{
Examples of one-dimensional photonic lattices modulated
by the third beam with the transverse wave number $k_{3x}$: (a) $k_{3x} = 0$, (b)
$k_{3x} = 0.8k_{12x}$, and (c) $k_{3x} = 1.3 k_{12x}$. Insets show the wave
vectors of two input beams which form the lattice, and the wave
vector of the third beam (red, dashed). Parameters are $A_{12} =
0.25$, $A_3 = 0.66 A_{12}$ and the propagation length is $L=50~$mm.
}

We note that propagation of broad solitons in such lattices was
discussed recently~\cite{Kartashov:physics/0502070:ARXIV} under
the conditions when weak longitudinal modulation acts on solitons
as an effective potential. In contrast, we show that the behavior
of strongly localized solitons is {\em dramatically different},
resulting in, for example, resonant soliton revivals for the lattice of Fig.~\rpict{potentials}(a), or a
sharp binary switching transition for deep asymmetric lattice
modulations [Figs.~\rpict{potentials}(b,c)]. These results are
not related to the effect of diffraction management earlier
discussed in Ref.~\cite{Eisenberg:2000-1863:PRL}, and they were
not reported in any of the earlier studies of the modulated
discrete systems~\cite{Ablowitz:2001-254102:PRL,
Peschel:2002-544:JOSB}.

\section{Binary soliton steering}

Propagation of an optical beam in an one-dimensional
optically-induced lattice can be described by a parabolic equation
for the normalized beam envelope $E(x,z)$,
\begin{equation} \leqt{nls}
   i \frac{\partial E}{\partial z}
   + D \frac{\partial^2 E}{\partial x^2}
   + {\cal F}(x,z,|E|^2) E= 0,
\end{equation}
where $x$ and $z$ are the transverse and propagation coordinates
normalized to the characteristic values $x_s$ and $z_s$,
respectively, $D = z_s \lambda / (4 \pi n_0 x_s^2)$ is the beam
diffraction coefficient, $n_0$ is the average refractive index of
the medium, and $\lambda$ is the wavelength in vacuum. The induced
change of the refractive index in a photorefractive crystal
is~\cite{Efremidis:2002-46602:PRE, Fleischer:2003-23902:PRL, Neshev:2003-710:OL, Sukhorukov:2004-93901:PRL}: ${\cal F}(
x, z, |E|^2) = - \gamma (I_b + I_p(x, z) + |E|^2)^{-1}$, where $I_b$
is the constant dark irradiance, $I_p(x, z)$ is the interference
pattern which induces modulations of the refractive index,
and $\gamma$ is a nonlinear coefficient proportional
to the applied DC field. In the numerical simulations presented
below we use the parameters which are typical for the experimental
conditions with optically-induced lattices created in
photorefractive crystals~\cite{Sukhorukov:2004-93901:PRL}:
$\lambda=0.532\mu$m, $n_0=2.35$, $x_s=1\mu$m, $z_s=1$mm, $I_b=1$, $\gamma=9.45$, 
and the transverse period of the lattice in the absence of modulation is $d=15\mu$m.

\pict{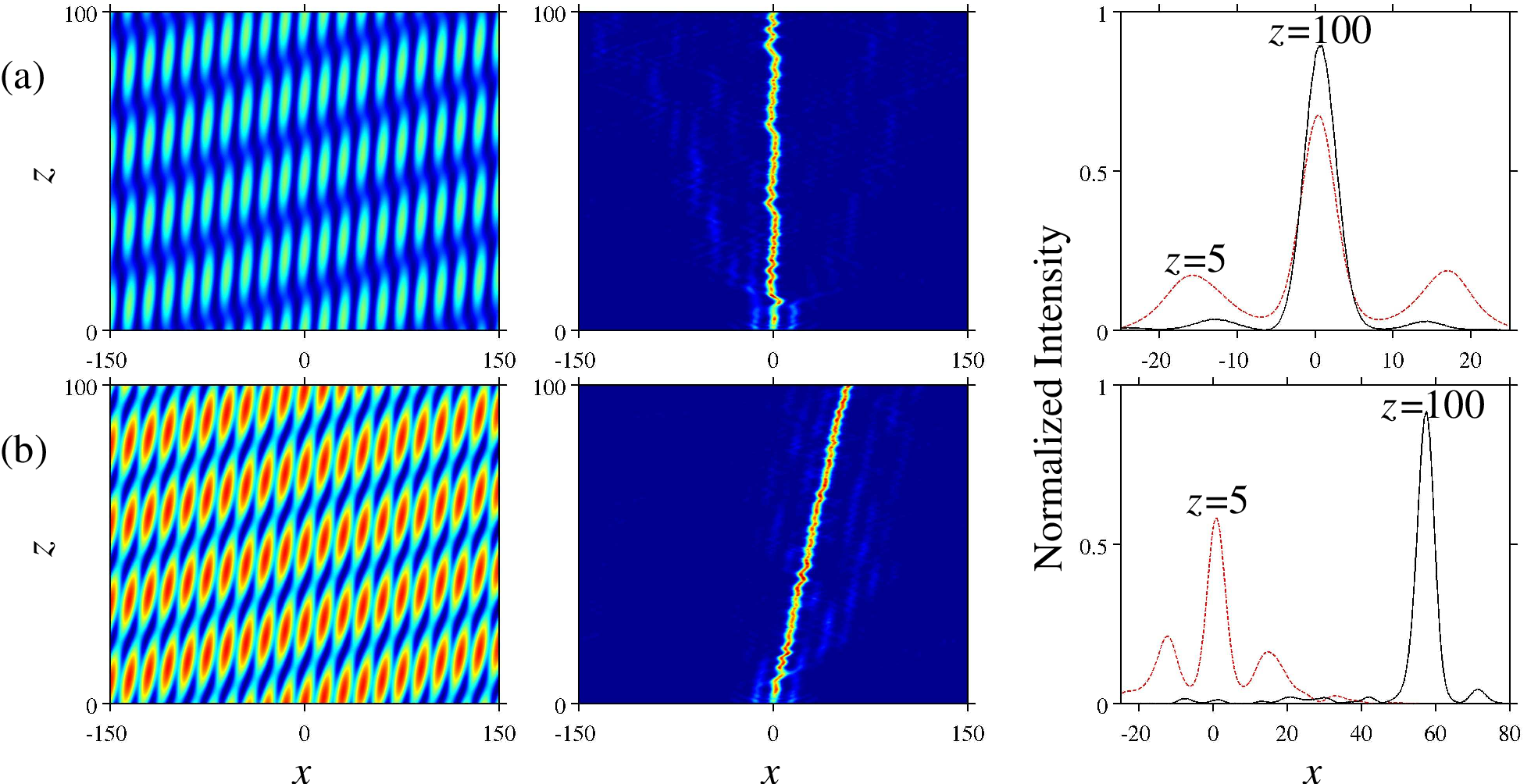}{steering}{
(1.8MB) All-optical steering of spatial optical solitons controlled by the amplitude of the third lattice beam with inclination $k_{3x} = 1.15 k_{12x}$: (a)~straight ($A_3 = 0.62 A_{12}$) and (b)~tilted ($A_3 = 2.02 A_{12}$) propagation. Left: profiles of optically-induced lattices. Middle: evolution of beam intensities along the propagation direction. 
Right: soliton profiles at the input (dashed) and output (solid).
Animation shows the soliton dynamics as the modulation depth increases from zero ($A_3=0$) to a higher value ($A_3=2.8
A_{12}$). Parameters are $A_{12}=0.25$, $A_{\rm in}=0.5$, input
beam position $x_0=0$, angle $k_{0x}=0$ and width $w=25\mu$m, and the total propagation length is $L=100~$mm.}

Photorefractive crystals exhibit a very strong electro-optic anisotropy, e.g. in SBN:75 the electro-optic coefficient for  extraordinary polarized waves is more than 20 times higher than the electro-optic coefficient for ordinary polarized waves~\cite{Fleischer:2003-147:NAT}. Thus, the lattice-writing beams polarized orthogonal to the {\em c}-axis of the crystal satisfy the same evolution Eq.~\reqt{nls}, but without the last term which almost vanishes since the effective nonlinear coefficient is very small~\cite{Efremidis:2002-46602:PRE}, while extraordinary polarized beam will experience a highly nonlinear evolution. 
Then, each of the broad lattice beams propagates independently, and it can be presented as a
linear plane-wave solution in the form $E_{\rm lattice} = A \exp(i
\beta z + i k_x x)$, where $k_x$ is the transverse wavenumber
proportional to the inclination angle, and the propagation
constant $\beta = - D k_x^2$ defines the longitudinal wavevector
component $k_z$. The value of diffraction coefficient $D$ can be controlled by varying the wavelength of lattice beams, and also depends on the crystal anisotropy. We will analyze in detail the case when the effective diffraction coefficients for the probe and lattice beams are the same, which also allows us to perform a comparison with the results of Ref.~\cite{Kartashov:physics/0502070:ARXIV}. Specifically, we consider a lattice induced by three
interfering waves~\cite{Kartashov:physics/0502070:ARXIV}: (i)~two
waves with equal amplitudes $A_{12}$ and opposite inclination
angles, with the corresponding wavenumbers $k_{12x}$ and $-k_{12x}$, and
(ii)~an additional third wave with amplitude $A_3$ and wavenumber
$k_{3x}$. Then, the optical lattice is defined through the wave
interference pattern $I_p(x,z)=|A_L|^2$, where
\begin{equation}
\leqt{lattice}
   A_L = A_3 \exp( i \beta_3 z + i k_{3x} x )
   + 2 A_{12} \exp( i \beta_{12} z ) \cos( k_{12x} x).
\end{equation}

It follows that additional beam (with $k_{3x} \ne k_{12x}$) always leads
to the lattice modulation both in the transverse and longitudinal
directions. We show examples of modulated lattices in
Figs.~\rpict{potentials}(a-c) corresponding to the same wave
amplitudes but different inclinations of the third beam (defined
by $k_{3x}$) as indicated in the insets. We see that for $k_{3x}=0$
[Fig.~\rpict{potentials}(a)] the lattice profile in the
transverse cross-section becomes double-periodic corresponding to
an alternating sequence of deeper and shallower potential wells
resembling a binary superlattice~\cite{Sukhorukov:2002-2112:OL},
however its configuration is periodically inverted due to
modulations in the longitudinal direction along $z$. On the other
hand, when $k_{3x} \simeq k_{12x}$, the lattice is slowly modulated in
both spatial directions and the left-right reflection symmetry is
removed~\cite{Kartashov:physics/0502070:ARXIV}, see
Figs.~\rpict{potentials}(b)~and~(c).

\pict{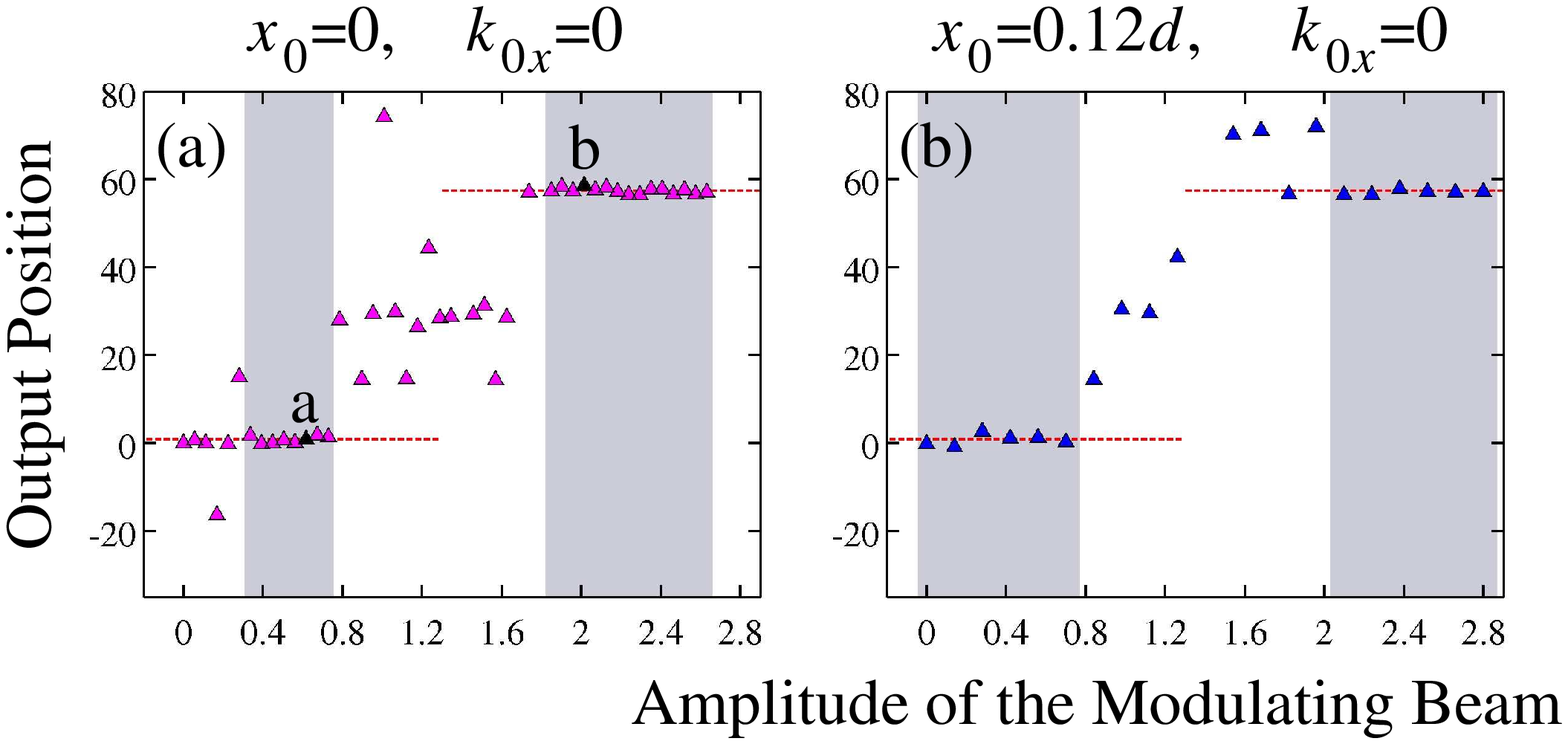}{transmission}{
Output soliton position vs. the modulating beam amplitude for different positions and angles of the input Gaussian beam. 
In (a)~marked points 'a', 'b' correspond to
the solitons shown in Fig.~\rpict{steering}(a) and
Fig.~\rpict{steering}(b), respectively.
Shadings mark stable regions. 
Parameters are the same as in Fig.~\rpict{steering}.
}

First, we consider the soliton dynamics in {\em asymmetric
lattices} with $k_{3x} \simeq k_{12x}$, and demonstrate the possibility
of {\em binary steering of strongly localized solitons}, where the
soliton propagates in one of two allowed directions when the
amplitude $A_3$ is in one of the two stable regions. The origin of
this soliton switching effect is fundamentally different from
dragging of broad solitons reported
earlier~\cite{Kartashov:physics/0502070:ARXIV} which is almost
directly proportional to the third beam amplitude $A_3$.

We perform numerical modelling of Eq.~\reqt{nls} to study
generation of a strongly localized lattice soliton by an input {\em Gaussian
beam},
\begin{math}
  E_{\rm in} = A_{\rm in} \exp \left\{ - [ (x - x_0) / w ]^2 + ik_{0x}(x - x_0)\right\}
\end{math}
which is incident on the crystal at normal angle (i.e. $k_{0x}=0$) and has extra-ordinary linear polarization.
When the amplitude of the third wave $A_3$ is relatively small,
the generated soliton starts moving between the neighboring lattice sites, as shown in the animation. As the amplitude $A_3$ of the modulating beam increases, at certain point strongly localized soliton becomes locked at a particular lattice site, and it propagates {\em straight} along the lattice [see Fig.~\rpict{steering}(a) and
Fig.~\rpict{transmission}(a)], similar to the case of homogeneous structures without longitudinal modulation~\cite{Morandotti:1999-2726:PRL}. We find that this is the first stable propagation regime which is not sensitive to small variations of the input angle and position [see Figs.~\rpict{transmission}(b,c)].

When the amplitude $A_3$ grows further, this leads not only to the increase in the modulation depth of the refractive index, but also to the rotation of the lattice high-index sites. This rotation causes the change in the topology of the
modulated optical lattice and in some interval of the modulation
depth there exists no continuous connectivity between the
high-index lattice sites (see animation in
Fig.~\rpict{steering}). In this regime the soliton propagation
can become highly irregular resembling a regime of random walks,
and the soliton can even be completely destroyed by the lattice
modulation.

At a certain value of $A_3$, the rotation of the lattice sites
experience saturation, the connectivity between the sites
reappear (but now it is diagonal in contrast to the first stable region where it was vertical),
and the soliton starts to move {\em across the lattice}
propagating in the direction determined by the angles of the
lattice waves and independent of the value of the modulation
amplitude $A_3$ [see Fig.~\rpict{steering}(b) and
Fig.~\rpict{transmission}(a)]. This is the second stable
propagation regime not sensitive to small variations in the input
conditions [see Figs.~\rpict{transmission}(b,c)]. At very high
values of the modulation amplitude $A_3$ the soliton do not form due to nonlinearity saturation.

\pict[0.68]{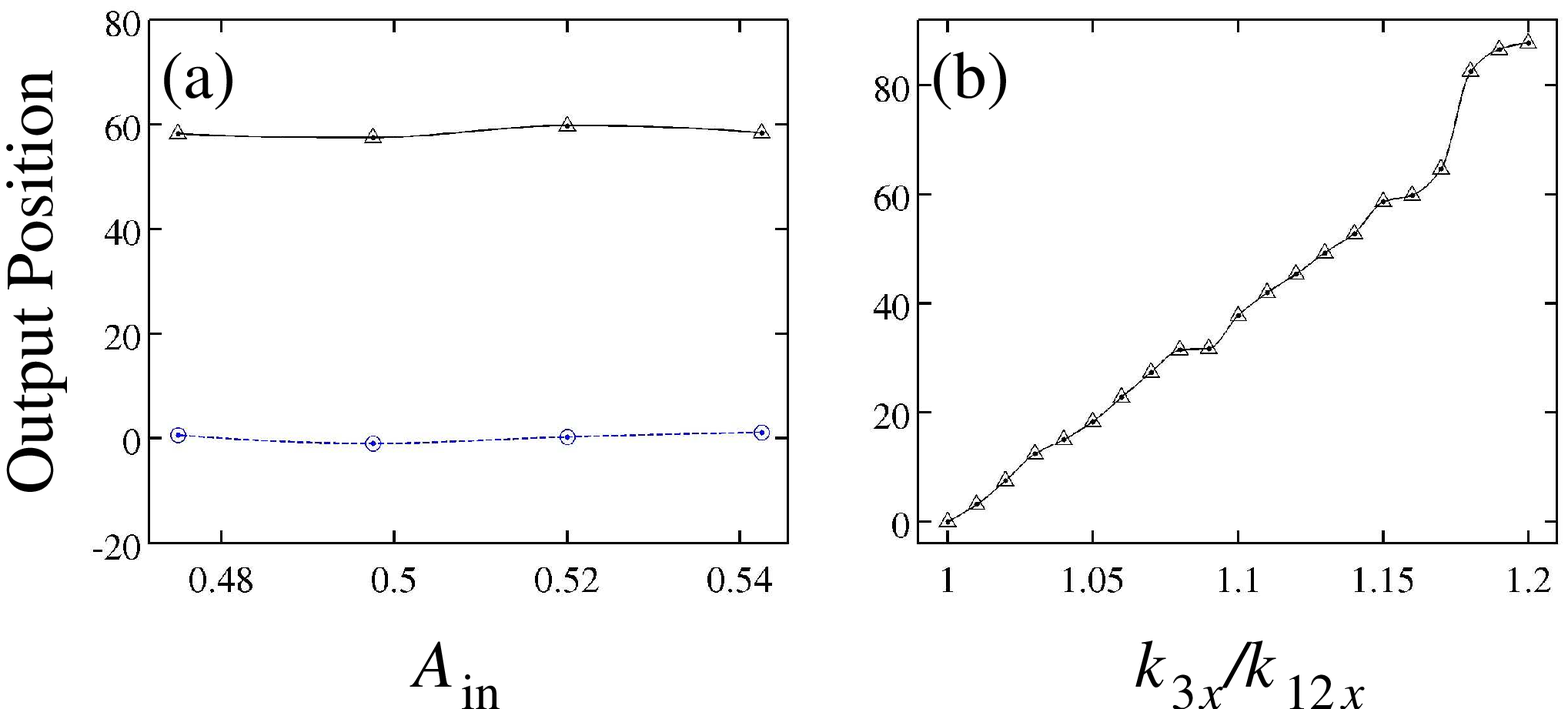}{dependences}{
Output soliton position vs. (a)~the amplitude of the input Gaussian beam and (b)~the angle between the modulating beam and the lattice-forming beams, defined by the ratio $k_{3x}/k_{12x}$. Dashed line
and circles correspond to Fig.~\rpict{steering}(a), solid lines
and triangles~-- to Fig.~\rpict{steering}(b).
Parameters are the same as in Fig.~\rpict{steering}.}

We can summarize that such {\em binary soliton steering} occurs due to
the substantial change in the geometry of the optical lattice,
where the connectivity between high-index lattice sites changes
from vertical to diagonal through a disconnected state when we increase the
amplitude of the third modulating wave, as illustrated by an animation in
Fig.~\rpict{steering}. Additionally, this binary soliton steering is found to be
insensitive to large variations of the soliton amplitude [see Fig.~\rpict{dependences}(a)], and the tilt of the soliton increases almost linearly with the difference between the angles of the modulating and lattice-forming beams [see Fig.~\rpict{dependences}(b)]. Such a behavior is completely different from the dynamics of broad solitons in weakly modulated lattices~\cite{Kartashov:physics/0502070:ARXIV}, which feel only spatially averaged, smoothed lattice potential. In contrast, behavior of strongly localized solitons is dominated by the fine geometrical structure of the lattice.

\section{Soliton revivals}

Next, we analyze the soliton dynamics in symmetric modulated lattices
when $k_{3x}=0$, as shown in Fig.~\rpict{potentials}(a). According
to the basic principles of holography, a beam which is incident on
the lattice at the normal angle (with $k_{0x}=0$) will  excite resonantly
the waves corresponding to other lattice-writing beams with
the transverse wavenumbers $\pm k_{12x}$, which will then be converted
back to the original wave. Numerical simulations indeed
demonstrate that the spectrum of low-amplitude beam is
{\em modulated periodically} as it spreads due to linear diffraction,
see Fig.~\rpict{explosion}(a).

\pict[0.7]{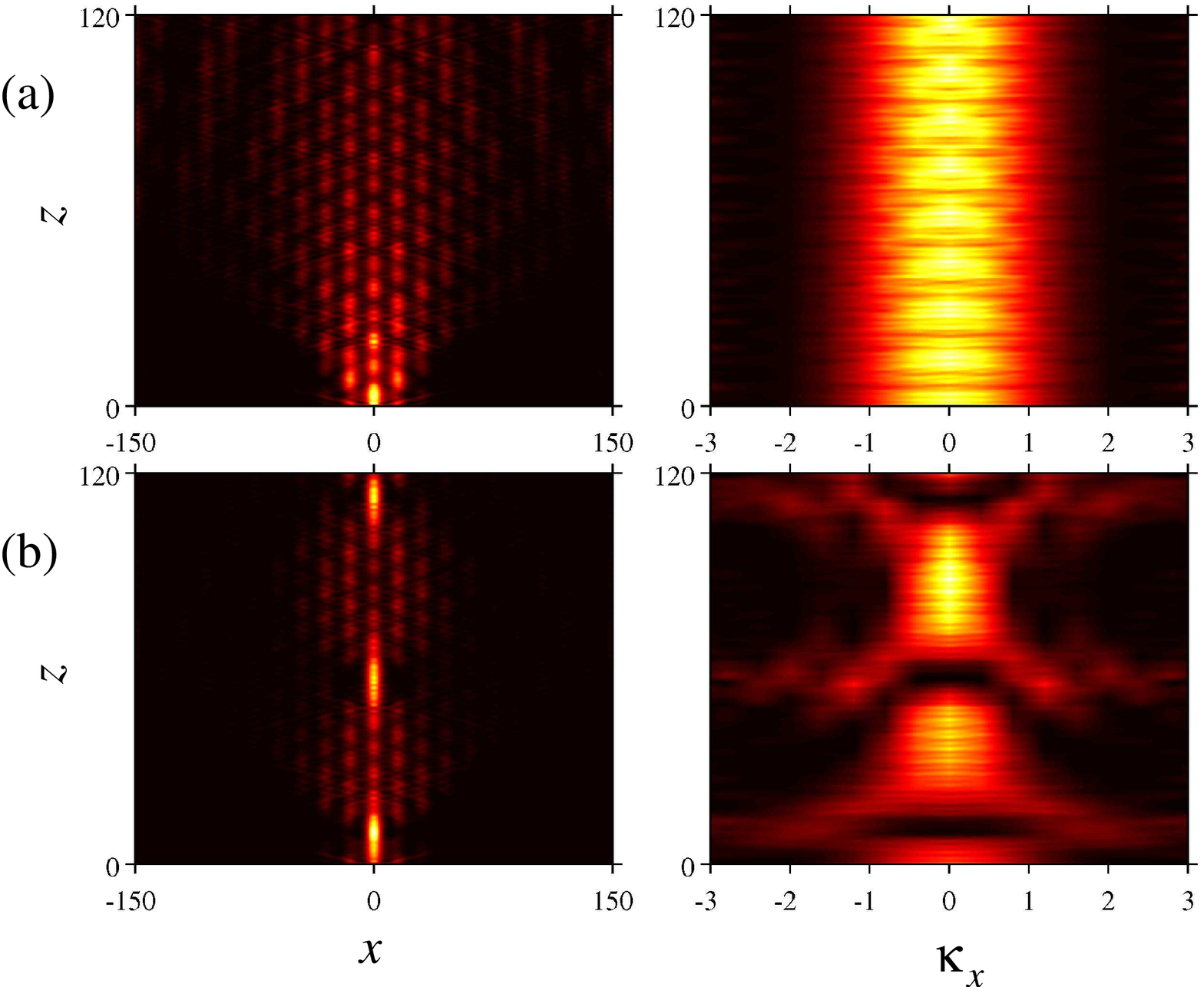}{explosion}{
Example of the resonant soliton revivals in the
modulated lattices: (a) linear diffraction at low power ($A_{\rm
in}=0.02$), (b) revivals and periodic transformations of the
soliton in the nonlinear regime ($A_{\rm in}=0.2$). Variation of 
the intensity (left) and spatial Fourier spectrum (right) of
the input Gaussian beam along the propagation direction are shown
(spatial frequency $\kappa_x$ is normalized to $k_{12x}/4$). Parameters are $A_{12}=0.25$, $A_3 =
0.2 A_{12}$, $x_0=0$, $k_{0x}=0$, $w=25\mu$m and the total propagation length is
$L=120~$mm.}

However, even for a weak lattice modulation, the beam dynamics is
dramatically modified at higher amplitudes, and we observe a
sequence of {\em soliton revivals}, as shown in
Fig.~\rpict{explosion}(b). We identify three regimes of the
soliton propagation: (i)~self-focusing of the beam which spatial spectrum is centered
around the point $\kappa_x=0$, (ii)~transformation of the modes from $\kappa_x=0$ to larger
spatial frequencies, (iii)~the spectrum conversion back to the
region around the point $\kappa_x=0$, and again a periodic repetition of
this three-stage process. We note that the period of the soliton revivals {does not
coincide} with the period of the lattice modulation  underlying {\em a
key difference} with the case of the familiar diffraction-managed
solitons~\cite{Ablowitz:2001-254102:PRL}. In our case, there
exists a {continuous coupling and transformation between the
modes of the periodic lattice}. More detailed analysis of these results, and discussions of their connection to the effects of soliton internal modes~\cite{Peschel:2002-544:JOSB} or formation of multi-band breathers~\cite{Mandelik:2003-253902:PRL} will be presented in a separate study.

\section{Conclusions}

We have demonstrated novel effects for the soliton propagation in
modulated dynamic optically-induced photonic lattices created by
three interfering beams. We have shown the possibility of binary
switching for strongly localized solitons where the soliton can
propagate in one of two allowed directions when the amplitude of
the control beam is below or above a threshold associated with the transformation of
the lattice geometry and a respective change in the connectivity between lattice sites. Each
of these regimes is stable with respect to the system parameters,
in contrast to earlier considered steering of broad beams directly
proportional to the control wave amplitude. We have also
demonstrated novel regimes in the soliton dynamics under the
conditions of resonant wave mixing in a conservative system observed as a series of
periodic soliton revivals, which are not associated
with the effect of diffraction management.

\end{sloppy}
\end{document}